\documentclass[conference]{IEEEtran}
\IEEEoverridecommandlockouts
\usepackage{cite}
\usepackage{algorithmic}
\usepackage{graphicx}
\usepackage{textcomp}
\usepackage{graphicx}
\usepackage{array}
\usepackage{array}
\usepackage{tcolorbox}
\usepackage{booktabs}
\usepackage{placeins}
\usepackage{listings} 
\usepackage{xcolor}   
\usepackage{hyperref}
\usepackage{everypage}
\usepackage{tikz}

\AddEverypageHook{%
  \begin{tikzpicture}[remember picture,overlay]
    \node[anchor=north,yshift=-0.5cm] at (current page.north) {\small\textit{Accepted in the 23rd IEEE/ACIS International Conference on Software Engineering, Management and Applications (SERA 2025)}};
  \end{tikzpicture}%
}

\def\BibTeX{{\rm B\kern-.05em{\sc i\kern-.025em b}\kern-.08em
    T\kern-.1667em\lower.7ex\hbox{E}\kern-.125emX}}

\begin{document}

\title{Exposing Go’s Hidden Bugs: \\ A Novel Concolic Framework}
\IEEEpeerreviewmaketitle 

\author{
\IEEEauthorblockN{
Karolina Gorna\IEEEauthorrefmark{1}\IEEEauthorrefmark{2},
Nicolas Iooss\IEEEauthorrefmark{2},
Yannick Seurin\IEEEauthorrefmark{2},
Rida Khatoun\IEEEauthorrefmark{1}
}
\IEEEauthorblockA{\IEEEauthorrefmark{1}\textit{Telecom Paris}\\
name.surname@telecom-paris.fr}
\IEEEauthorblockA{\IEEEauthorrefmark{2}\textit{Ledger Donjon}\\
name.surname@ledger.com}
}

\maketitle

\begin{abstract}
The widespread adoption of the Go programming language \cite{go-community_go_2025} in infrastructure backends and blockchain projects has heightened the need for improved security measures. Established techniques such as unit testing, static analysis, and program fuzzing provide foundational protection mechanisms. Although symbolic execution tools have made significant contributions, opportunities remain to address the complexities of Go's runtime and concurrency model. In this work, we present \textit{Zorya}, a novel methodology leveraging concrete and symbolic (concolic) execution to evaluate Go programs comprehensively. By systematically exploring execution paths to uncover vulnerabilities beyond conventional testing, symbolic execution offers distinct advantages, and coupling it with concrete execution mitigates the path explosion problem. Our solution employs Ghidra's P-Code \cite{naus_formal_2023} as an intermediate representation (IR). This implementation detects runtime panics in the TinyGo compiler \cite{tinygo-org_tinygo_2019} and supports both generic and custom invariants. Furthermore, P-Code's generic IR nature enables analysis of programs written in other languages such as C. Future enhancements may include intelligent classification of concolic execution logs to identify vulnerability patterns.\\
\end{abstract}


\section{Introduction}
Go's rising popularity in cloud, infrastructure, and blockchain applications raises security concerns due to its unique runtime and concurrency features (goroutines, channels). Reports indicate over 66\% of Go modules contain vulnerabilities \cite{hu_empirical_2024}, highlighting the critical need for robust security analysis.  Challenges include unsafe memory operations, pointer misuse, complex error handling, and the unpredictable nature of garbage collection and goroutine management.  These factors, compounded by the intricacies of Go's concurrency model, raise the question of how to effectively identify and mitigate vulnerabilities in such a complex environment.

\textbf{Contributions.} To address these challenges, this paper proposes an innovative approach using concolic execution for the security verification of Go programs. Our framework leve\-rages P-Code as an intermediate representation to model with granularity Go’s execution semantics. 
This work improves security verification by:
\begin{itemize}
    \item Proposing a new approach for generating and parsing P-Code outside of Ghidra \cite{nsa_ghidra_2017} \cite{eagle_ghidra_2020}, including examples of test programs and their corresponding P-Code files.
    \item Proposing a novel concolic execution method to uncover common vulnerabilities in Go, and other languages with binaries convertible to P-Code, such as C.
    \item Implementing an open-source Proof-of-Concept of the concolic execution method and validating it with a custom dataset of binaries and encoded vulnerabilities.
\end{itemize}
This paper begins by overviewing techniques for securing Go code and their corresponding tools, followed by a comprehensive presentation of our contributions. It then details our proposed methodology, presents our evaluation results, reviews related work, and concludes with a summary of our contributions and potential future research directions.

\section{Background: Securing Go Code}

Security tools for Go address vulnerabilities using diverse methods, each with distinct strengths and limitations. Static analysis tools such as \textit{gosec} \cite{securego_gosec_2024}, \textit{go-vet} \cite{google_govet_2024}, \textit{staticcheck} \cite{google_staticcheck_2018}, and \textit{errcheck} \cite{kisiel_kisielkerrcheck_2024} perform analyses to detect common issues like unchecked errors and unsafe pointer usage. While these tools integrate well into continuous integration workflows, their semantic checks can miss deeper vulnerabilities.

Dependency-focused tools like \textit{Snyk} \cite{snyk_snykcli_2024} detect known vulnerabilities within third-party modules by continuously scanning dependency graphs against vulnerability databases, but they can overlook deeper application-specific logic errors.

In contrast, \textit{CodeQL} \cite{codeql_codeql_2021} provides a query-based static ana\-lysis framework supporting sophisticated data-flow analyses to detect complex vulnerabilities. Nevertheless, its effectiveness requires significant setup effort and query-writing expertise.

Dynamic analysis tools, notably \textit{go-fuzz} \cite{vyukov_dvyukovgo-fuzz_2024} and Google's \textit{gofuzz} \cite{google_googlegofuzz_2024}, use randomized inputs to uncover runtime bugs such as buffer overflows or panics. Although effective at revealing edge cases, their limited path exploration can restrict their capability to identify state-dependent logical errors.

Black-box frameworks like \textit{gopter} \cite{leanovate_leanovategopter_2024} enhance fuzz testing by enabling stateful property-based testing; however, their lack of white-box insight can limit their effectiveness.

Specialized tools like \textit{krf} \cite{trail-of-bits_trailofbitskrf_2024} and \textit{on-edge} \cite{trail-of-bits_trailofbits-edge_2023} specifically target vulnerabilities related to Go’s \textit{defer}, \textit{panic}, and \textit{recover}, but their narrow scope and infrequent updates reduce their effectiveness against evolving threats.

Existing tools address specific vulnerability classes but can remain insufficient for detecting complex flaws.

\section{Overview of the contributions}

\subsection{Core contributions}

\begin{figure}[htbp]
\centerline{\includegraphics[scale=0.39]{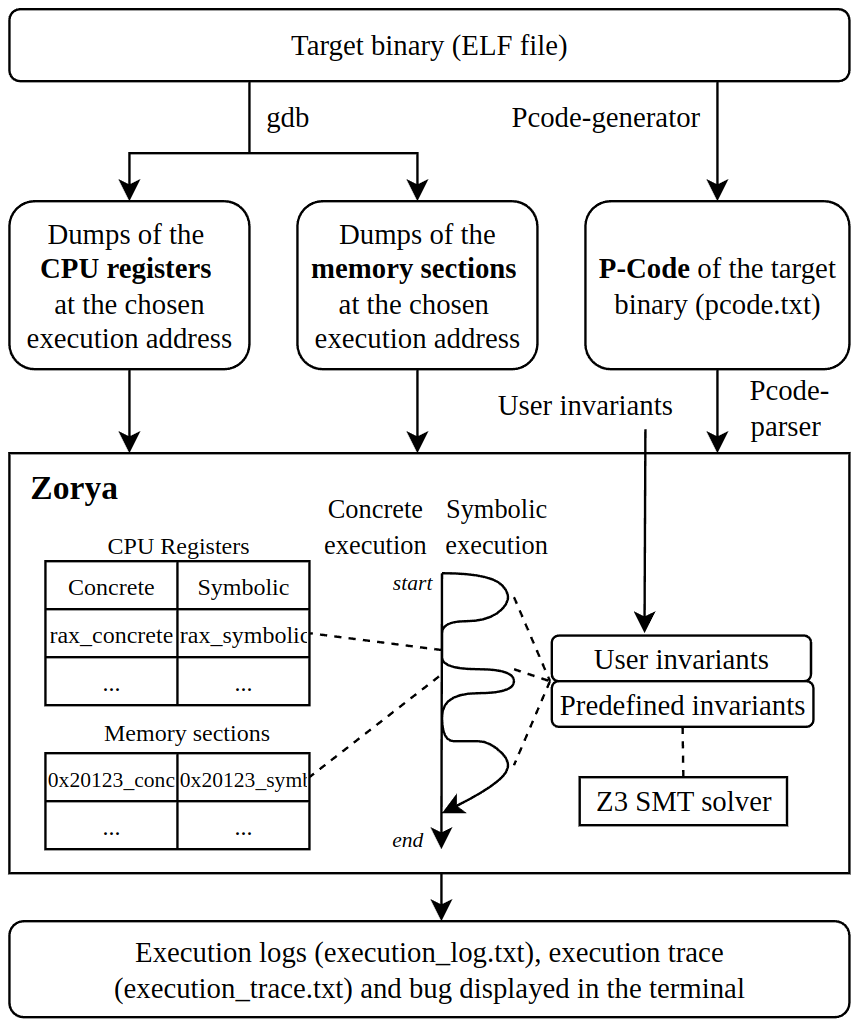}}
\caption{Overview of the contributions, including \textit{Zorya}, \textit{Pcode-generator} and \textit{Pcode-parser}.}
\label{fig}
\end{figure}

\textit{Zorya} is a concolic execution framework designed to detect logic-related bugs, language-specific vulnerabilities, and uncover new patterns of security issues, primarily in Go binaries. As illustrated in Fig.~\ref{fig}, the analysis begins by generating CPU registers and memory dumps using \textit{gdb} \cite{stallman_debugging_2018} at a user-specified address. \textit{Zorya} then loads these dumps to initialize execution from the given starting point, ensuring a realistic and accurate representation of the program state.

The core methodology involves translating binary code into raw P-Code, a low-level intermediate representation, which is subsequently parsed for precise execution path analysis. \textit{Zorya} focuses on key aspects such as targeting the sections containing executable code in studied binaries, supporting full binaries with runtime components, and accommodating shared libraries. Due to the lack of a complete Rust-based library for P-Code parsing, \textit{Pcode-parser} was implemented from scratch, improving upon incomplete alternatives like \textit{sleigh-rs} \cite{brandao_sleigh-rs_2024}.

\textit{Zorya}'s engine, implemented in Rust, uses the \textit{Z3 SMT solver} \cite{de_moura_z3_2008} and includes a state manager, a CPU state, a memory model, and a virtual file system. It emulates P-Code instructions (e.g. \texttt{handle\_int\_add}, \texttt{handle\_load}) to track the execution and detect vulnerabilities in the analyzed binaries. \textit{Zorya} supports both concrete and symbolic data types, x86-64 instructions and syscalls, and manages the program counter. Currently, \textit{Zorya} analyzes single-threaded Go programs compiled with TinyGo, with plans to address multithreading and goroutines in future work.

\subsection{Proof-of-Concept Functionalities}

\textit{Zorya}'s current Proof-of-Concept demonstrates several advanced functionalities essential to effective concolic execution analysis. It notably facilitates precise concolic handling of jump tables, which are specialized switch table constructs replacing traditional binary searches with direct jumps for consecutive numeric labels (see \texttt{jump\_table.json}). Furthermore, it can systematically identify and document cross-reference addresses leading directly to panic functions embedded within Go target binaries, considerably aiding in targeted vulnerability assessments (see \texttt{xref\_addresses.txt}). Additionally, \textit{Zorya} is proficient in translating dynamically loaded executable sections of shared libraries, such as \texttt{libc.so} and \texttt{ld-linux-x86-64.so}, into P-Code, providing robust analysis capabilities for dynamically linked binaries.

The current Proof-of-Concept implementation of \textit{Zorya} also generates comprehensive execution logs, recording step-by-step instruction-level details for thorough analysis (see \texttt{execution\_log.txt}). Additionally, \textit{Zorya} systematically captures the executed symbols, including functions and their arguments' values, facilitating the tracking and recording of the effective execution (see \texttt{execution\_trace.txt}). These detailed insights significantly enhance the ability to reconstruct execution paths, identify potential execution bottlenecks, and debug concolic execution processes.

\section{Methodology}
\label{sec:methodology}

\subsection{Concolic execution and Go analysis}
Symbolic execution treats program variables as symbolic variables to explore all possible execution paths, while concrete execution runs the program with actual input values. Dynamic symbolic execution (DSE) benefits from the efficiency and decidability of concrete execution and the stronger guarantees of symbolic execution. Concolic execution, is a specific type of DSE that uses concrete execution to drive symbolic execution, building symbolic representations while executing with concrete values. This approach helps in generating new concrete inputs (test cases) to maximize code coverage and find bugs in real-world software.
By running a program with real input values while simultaneously treating certain inputs as symbolic variables, concolic execution mitigates the exponential path growth encountered in purely symbolic approaches. This makes it more practical for analyzing large binaries, such as Go-Ethereum’s Geth, which has a size of 70 MB when unoptimized \cite{ethereum-foundation_go-ethereum_nodate}.
However, few symbolic or concolic execution tools can effectively analyze Go programs. This limitation arises primarily from their lack of support for multithreading and system calls, which are prevalent in Go's non-deterministic runtime.
Table~\ref{tab:symbolic_execution_tools} indicates that radius2 and MIASM are among the few tools offering basic compatibility with Go. \textit{Zorya} aims to provide the most adapted concolic execution framework for Go analysis.

\FloatBarrier 
\begin{table*}[tb]
\caption{Comparing \textit{Zorya} to existing Symbolic-Execution-Based Tools \\ (SE: Symbolic Execution / CE: Concrete Execution)}
\label{tab:symbolic_execution_tools}
\begin{center}
\renewcommand{\arraystretch}{1.5}
\begin{tabular}{|p{1.3cm}|p{0.8cm}|p{2.3cm}|p{1.6cm}|p{1.6cm}|p{3.5cm}|p{1.7cm}|p{2cm}|}
\hline
\textbf{Tool} & \textbf{Lan-guage} & \textbf{Method } & \textbf{Target} & \textbf{Intermediate Representation (IR)} & \textbf{IR and Architecture Adapted for Go (limited / moderate / advanced)} & \textbf{Integrated SMT Solver(s)} & \textbf{User Interface} \\
\hline
MAAT & C++ & SE using CE & Binaries & LLVM IR & No, gollvm is not maintained & Z3 & CLI only \\
\hline
Haybale & Rust & SE & Binaries & LLVM IR & No, gollvm is not maintained & Boolector & CLI only \\
\hline
Triton & C++ & SE using CE & Binaries & Custom IR & No, lacking support & Z3 & CLI and IDA Pro plugin (paid)\\
\hline
KLEE & C++ & SE & LLVM bitcode & LLVM IR & No, gollvm is not maintained & MetaSMT, STP and Z3 & CLI and web UI \\
\hline
Angr & Python & SE using CE + Binary Analysis & Binaries & VEX, P-Code & No, VEX not suitable for Go and no full support for P-Code & Z3, Boolector and CVC4 & CLI and GUI \\
\hline
SymSan & C++ & SE using CE & Binaries & LLVM IR & No, gollvm is not maintained & Z3 & CLI only \\
\hline
Fuzzolic & C & SE using Fuzzing & Binaries & QEMU-based & No, lacking support & Fuzzy-SAT & CLI only \\
\hline
DuckEEGO & Go & SE using CE & Source Code & Go AST & Yes, limited & Z3 & CLI only \\
\hline
Radius2 & Rust & SE + Taint Analysis & Binaries & ESIL & Yes, limited & Z3, Boolector & CLI only \\
\hline
MIASM & Python & SE using CE + Binary Analysis & Binaries & Custom IR & Yes, limited & Z3, CVC4 & CLI and IDA Pro plugin (paid)\\
\hline
Zorya & Rust & CE and SE + Binary Analysis & Binaries & P-Code and emulation & Yes, moderate & Z3 & CLI and GUI in Ghidra (free) \\
\hline
\end{tabular}
\end{center}
\end{table*}
\FloatBarrier

\subsection{P-Code as Intermediate Representation}
P-Code, Ghidra’s intermediate representation language, offers significant advantages for this research due to its integration with the robust disassembly framework released by the NSA in 2017. Compared to alternatives such as LLVM IR (with the not fully maintained \textit{gollvm} compiler \cite{go-community_gollvm_2017}), VEX \cite{nethercote_valgrind_2007}, BIL \cite{brumley_bap_2011}, or REIL \cite{li_dynamic_2013}, P-Code provides a granular abstraction that combines low-level semantic detail with structural analyzability, and is tightly coupled with Ghidra’s lifting pipeline. Other formats, such as WASM, are rarely used in backends or blockchain clients, limiting their relevance in this context. Similarly, performing symbolic execution directly on x86 machine code lacks the abstraction and maintainability offered by an IR.

While P-Code itself refers to the low-level IR generated by Ghidra’s Sleigh specifications, the decompiler constructs a higher-level representation on top of it, exposing control-flow, data-flow, and recovered variables. Although not formally a separate IR, this higher-level view is often used for source-like analysis \cite{10.1145/3597926.3598039}. In this work, we make the novel choice to operate directly on low-level P-Code, as its finer granularity preserves detailed instruction semantics—an essential property for precise symbolic reasoning on optimized or semantically intricate binaries.

In addition, Ghidra’s dedicated Go lifter \cite{nsa_ghidras_2025} improves the accuracy of analysis by preserving language-specific features such as runtime metadata, goroutines, and channels. Custom Java scripts further extend Ghidra’s capabilities for Go-specific inspection and analysis \cite{trellix_advanced_research_center_ghidrascripts_2024}.

\subsection{Bugs detection} Several strategies, optimized for Go binaries, can be implemented in \textit{Zorya} to detect bugs effectively. The first strategy (S1) employs concrete execution combined with a flag-raising mechanism. This mechanism monitors the execution flow and triggers a signal when the program approaches the invocation of a panic function, as identified through the symbol list. This indicates that an error-inducing branch has been encountered. The second strategy (S2) integrates both concrete and symbolic execution (concolic execution). This approach utilizes a Z3-based symbolic invariant defined as: \textit{"The program counter must never point to an address that is a cross-reference to a panic function."} This invariant ensures systematic exploration of paths to verify whether any execution violates the constraint, identifying potential vulnerabilities. For example, this invariant prevents crashes from nil pointer dereferences, a common cause of denial-of-service vulnerabilities. The third strategy (S3) focuses on targeted concolic analysis of specific functions. \textit{Zorya} initiates execution at the function's address, with its arguments populated by symbolic variables and randomized concrete values. This hybrid approach allows for guided execution while verifying the satisfiability of custom invariants.

These strategies are designed to be complementary. By default, at each execution of \textit{Zorya} on a binary, strategy (S1) is enabled. The analysis can then proceed by examining the entire binary starting from the \textit{start} or \textit{main} address (strategy (S2)), or by initiating execution at the address of a specific function (strategy (S3)).

\section{Running example}

Listing~\ref{lst:nilMapPanic} presents a minimal Go code where assigning to a nil map triggers a runtime panic. Detecting this bug with \textit{Zorya} involves the following steps. First, the code is compiled with TinyGo, and the resulting binary is translated to P-Code. Next, \textit{Zorya} executes the P-Code, starting from the \textit{main.main} address. During execution, \textit{Zorya} updates its concolic state at each P-Code instruction. 

Since the bug is embedded in the code, the first bug detection strategy (S1) described in \autoref{sec:methodology} is employed. This strategy identifies the runtime panic at address \texttt{0x2034c5}. Specifically, when the program counter reaches this address, \textit{Zorya} raises a flag, halts the analysis, and reports an attempt to \textit{"add an entry to a nil map."} This example, along with detailed evaluation results, is available in the Zorya-evaluation repository*.

\begin{lstlisting}[language=Go, caption={Example of a Go program attempting to assign to a nil map.}, label={lst:nilMapPanic}]
package main

func nilMapPanic() {
    var m map[string]int // nil map
    m["key"] = 42        // triggers panic
}

func main() {
    nilMapPanic()
}
\end{lstlisting}

\section{Evaluation}
In this section, we evaluate our approach based on the following research questions (RQ):
\begin{itemize}
    \item \textbf{RQ1}: Could P-Code be generated and used outside of Ghidra's framework?
    \item \textbf{RQ2}: How does our method compare to existing symbolic or concolic execution approaches in terms of Go bugs detection?
    \item \textbf{RQ3}: Can our method be used for other languages bugs detection?
\end{itemize}
All our experiments are conducted on a 64-bit machine with Linux, Ghidra 11.0.3, TinyGo v0.33.0
and gcc v11.4.0.

\vspace{0.2cm}

\textbf{RQ1: Generating and Using P-Code Outside Ghidra's Framework.}

Ghidra’s lack of an API for external P-Code generation required developing custom tools. Using its Java classes, we built the \textit{Pcode-generator} to extract and save low- or high-level P-Code. Key challenges included symbol mapping from \texttt{.text} and \texttt{.rodata} sections and implementing a low-level P-Code parser aligned with Ghidra's x86-64 specifications. Results (Table~\ref{tab:comparison}) show high accuracy, with no false positives for Go binaries and minimal issues for C binaries, mainly caused by complex structures. File generation is efficient, confirming the feasibility of external P-Code use for Go and C program analysis. Complex structures or \texttt{libc} usage did not affect false positives. File generation takes only a few seconds, depending on binary size, ensuring efficient processing.

\begin{table}[h]
    \centering
    \caption{Measuring accuracy of the generation of P-Code\\ according to binary source code language}
    \begin{tabular}{lcccc}
        \toprule
        \textbf{Binary Type} & \textbf{True Positive} & \textbf{False Positive} & \textbf{Total} \\
        \midrule
        Go & 10 & 0 & \textbf{10} \\
        C & 9 & 1 & \textbf{10} \\
        \bottomrule
    \end{tabular}
    \label{tab:comparison}
\end{table}
\vspace{0.2cm}

\noindent
\colorbox{gray!10}{%
    \parbox{0.47\textwidth}{
        \textbf{Finding 1:} P-Code, generated in low or high representation, can be used outside Ghidra for Go and C programs, though limitations arise for more complex structures.
    }
}
\vspace{0.1cm}

\textbf{RQ2: Comparison with Existing Approaches.} 

Table~\ref{tab:comparison2} summarizes the detection results for common Go vulnerabilities associated with TinyGo runtime panics. The assessment was conducted using a benchmark of small Proof-of-Concept programs replicating widely known Go vulnerabilities, as detailed in the Zorya Evaluation repository*. These vulnerabilities include critical issues described in the TinyGo documentation \cite{tinygo-org_tinygos_2025}, such as nil pointer dereference, out-of-bounds index access, nil map assignments, excessive channel creation, and negative bit shifts. However, the three symbolic/concolic execution tools evaluated—\textit{DuckEEGO}, \textit{Radius2}, and \textit{MIASM}—failed to detect any of these vulnerabilities. 

\textit{DuckEEGO}, originally developed for Go 1.10, faces compatibility challenges due to significant changes introduced in modern Go versions. The adoption of Go modules mandates explicit \texttt{go.mod} files, breaking previous GOPATH-based dependency resolution. Additionally, stricter enforcement in the \texttt{reflect} package results in failures during dynamic method resolution, necessitating explicit pointer receivers and enhanced error handling. Furthermore, \texttt{go build} no longer supports GOPATH-only projects, requiring a module-based compilation approach. While these issues were mitigated by manually initializing Go modules, adding \texttt{replace} directives, and refining method lookups, these adaptations were insufficient for \textit{DuckEEGO} to detect any of the vulnerabilities in our benchmark. 

\textit{Radius2} was tested on Go binaries, but in all cases, the analysis terminated unexpectedly at arbitrary points without providing information on execution status or conclusions. This lack of transparency hindered its practical applicability for Go vulnerability detection.

\textit{MIASM} requires a Python configuration file specifying the strategy for identifying target vulnerabilities. However, its detection mechanism assumes that the bug is actively triggered during execution. In the case of nil pointer dereference, \textit{MIASM} expects to dereference the pointer to observe the fault. Yet, Go binaries are compiled to redirect such operations to a panic routine instead of executing the faulty instruction directly. Consequently, \textit{MIASM} fails to detect these vulnerabilities as it has not been adapted to account for Go’s panic handling mechanisms.

In contrast, \textit{Zorya} successfully identified all vulnerabilities without false positives. Its effectiveness is attributed to its reliance on concrete execution and the first detection stra\-tegy (S1) described in \autoref{sec:methodology}, which flags panics when specific program counter values are reached. Additionally, \textit{Zorya}'s detection workflow is highly efficient, completing analyses in under a minute. Its simple interface (\texttt{zorya <path/to/bin>}) supports interactive mode, allowing users to select the starting address and define custom invariants.
\begin{table}[h]
    \centering
    \caption{Comparison of Go runtime bug detection across different methods (Detected (D) / Not Detected (ND))}
    \renewcommand{\arraystretch}{1.2} 
    \setlength{\tabcolsep}{4pt} 
    \begin{tabular}{>{\raggedright}p{3cm}cccc} 
        \toprule
        \textbf{Go Bugs} & \textbf{DuckEEGO} & \textbf{radius2} & \textbf{MIASM} & \textbf{Zorya} \\
        \midrule
        Nil Pointer Dereference & ND & ND & ND & \textbf{D} \\
        Index Out Of Range & ND & ND & ND & \textbf{D} \\
        Nil Map Assignment & ND & ND & ND & \textbf{D} \\
        Too Large Channel Creation & ND & ND & ND & \textbf{D} \\
        Negative Shift & ND & ND & ND & \textbf{D} \\
        \bottomrule
    \end{tabular}
    \label{tab:comparison2}
\end{table}

\vspace{0.2cm}

\noindent
\colorbox{gray!10}{%
    \parbox{0.47\textwidth}{
        \textbf{Finding 2:} Our method demonstrates improved performance over other symbolic execution approaches in the detection of common runtime bugs in Go using the TinyGo compiler.
    }
}
\vspace{0.1cm}

\textbf{RQ3: Extending \textit{Zorya} to bug detection in C.}
To evaluate \textit{Zorya}’s ability to analyze binaries beyond Go, we tested it on three C programs featuring common vulnerabilities. By defining relevant invariants, \textit{Zorya} successfully identified all issues. The first, a \textit{null dereference}, was detected by checking during STORE and LOAD operations whether the pointer was null. The second, \textit{misaligned memory}, was identified by verifying if the Euclidean division of the LOAD address by the loaded size yielded zero; otherwise, the memory was misaligned. The third, \textit{use of an uninitialized variable}, was detected by confirming that any loaded address had been previously stored in memory. As this approach incorporates concrete execution, it avoids false positives while maintaining the efficiency and simplicity of \textit{Zorya}’s commands demonstrated in RQ2.

\vspace{0.3cm}

\noindent
\colorbox{gray!10}{%
    \parbox{0.47\textwidth}{
        \textbf{Finding 3:} Our method can be used on C binaries to detect null pointer dereferences, misaligned memory bugs and the usage of simple uninitialized variables.
    }
}
\newline

\section{Discussion and Vision}

Currently, \textit{Zorya} identifies vulnerabilities related to TinyGo compiler panics in a single-threaded context. It must be extended to simulate multi-threaded execution to support programs built with the Go compiler and detect potential race conditions. On the symbolic side, \textit{Zorya} still requires comprehensive evaluation, particularly in refining detection strategies (S2) and (S3) described earlier. The current constraints stem from a limited symbolic exploration depth, which hinders the discovery of complex paths leading to panics. For C binaries, \textit{Zorya} performs basic checks, such as preventing invalid pointer dereferences, but additional invariants and analysis techniques could be integrated. Moreover, advanced strategies, such as intelligent classification of concolic variables, may improve its ability to detect vulnerability patterns.

\section{Related Work}
Table~\ref{tab:symbolic_execution_tools} presents a detailed comparison of prominent symbolic execution tools, emphasizing their implementation languages, intermediate representation (IR) methods, and their specific compatibility or limitations with the Go language ecosystem. Tools such as \textit{MAAT} \cite{trail_of_bits_maat_2024}, \textit{Haybale} \cite{plsyssec_haybale_2024}, and \textit{SymSan} \cite{r-fuzz_symsan_2024} are primarily developed in C++ and Rust, utilizing LLVM Intermediate Representation (LLVM IR) to achieve symbolic execution at a low abstraction level. However, these tools inherently face compatibility challenges when directly analyzing Go binaries due to the \textit{gollvm} compiler lacking of many functionalities \cite{go-community_gollvm_2017}.

In contrast, more versatile symbolic execution platforms such as \textit{Angr} \cite{wang_angr_2017} and \textit{MIASM} \cite{miasm_miasm_2015}, despite their robustness in different binary formats and architectures, exhibit IR compatibility issues with Go binaries. This incompatibility predominantly arises because Angr relies heavily on VEX IR and a P-Code emulation, which encounters difficulty accurately modeling Go-specific runtime structures, garbage collection routines, and goroutine management. Similarly, MIASM, leveraging Python-based modular architectures, experiences limited efficiency when dealing with Go's statically compiled binaries and internal abstractions, necessitating additional translation or adaptation layers.

\textit{Radius2} \cite{aemmitt-ns_radius2_2024}, built on Radare2 \cite{radare2-org_radare2_2024}, uses the ESIL intermediate language within a flexible, command-driven framework. However, ESIL's coarse abstraction often requires significant customization to support precise concolic execution, particularly for Go's concurrency and complex memory model.

Additionally, Ghidra plugins like \textit{GhiHorn} \cite{cert_coordination_center_ghihorn_2021} and \textit{CERT Kaiju} \cite{cert_coordination_center_cert_2024}, though not explicitly detailed in Table~\ref{tab:symbolic_execution_tools}, were critically evaluated for their capabilities in path-sensitive analysis and handling SMT (Satisfiability Modulo Theories) constraints directly within Ghidra's interactive interface. These plugins are inherently limited by their tight coupling with Ghidra's user-driven workflows and Java-based architecture, constraining their scalability and the degree of automated symbolic reasoning achievable with Go binaries.

Lastly, \textit{DuckEEGO} \cite{shao_duckee_2018} is a source-level concolic execution framework for Go that operates on the abstract syntax tree (AST) prior to compilation. It supports only basic types—\texttt{int}, \texttt{bool}, and \texttt{map[int]int}—and lacks support for strings, structs, floating-point numbers, external libraries, and runtime functions. It also does not handle multiple return values, goroutines, or syscalls, making it unsuitable for concurrent or system-level analysis. As it transforms source code rather than binaries, it cannot be applied to closed-source programs or Go runtime internals. 

\section{Conclusion}

Our methodology and associated implementation demonstrate that using Ghidra's P-Code as an IR effectively identifies vulnerabilities in single-threaded Go and C binaries, particularly revealing issues such as TinyGo compiler panics or unsafe pointer dereferences. However, further improvements are necessary, especially in the symbolic execution phase, to refine strategies for systematically detecting vulnerabilities through deeper path exploration and more precise invariants. Additional work should focus on expanding \textit{Zorya}’s capabilities to model multi-threaded Go programs, enabling the identification and verification of race conditions and concurrent execution issues. Moreover, future work may include smart classification of concolic execution logs to identify vulnerability patterns.

\vspace{0.3cm}

\noindent
\colorbox{gray!10}{%
    \parbox{0.47\textwidth}{
        \textbf{\textit{Zorya}:} https://github.com/Ledger-Donjon/zorya\\
        \textbf{\textit{Pcode-generator}:}\\ https://github.com/Ledger-Donjon/pcode-generator\\
        \textbf{\textit{Pcode-parser}:}\\
        https://github.com/Ledger-Donjon/pcode-parser\\
        \textbf{*Zorya-evaluation:}\\ https://github.com/Ledger-Donjon/zorya-evaluation
    }
}

\section*{Acknowledgment}
For valuable discussions that contributed to this work, we would like to thank Patrick Ventuzelo, Charles Christen and the Ledger Donjon team, Prof. Martin Monperrus and the KTH ASSERT team, Prof. Roberto Guanciale, Dr. Robin David, Rubens Brandão and Dr. Josselin Feist. GPT-4 was used to refine the paper's writing style.

\bibliographystyle{IEEEtran}

\end{document}